\documentclass[aps,prl,twocolumn,showpacs,preprintnumbers,letterpaper]{revtex4}
\usepackage{amssymb}

\usepackage{graphicx}
\usepackage{amsmath}
\usepackage{times}
\usepackage{color}
\usepackage{subfigure}
\usepackage{hyperref}

\begin{document}

\title{Lattice Induced Resonances in One Dimensional Bosonic Systems}
\author{Javier von Stecher$^1$, Victor Gurarie$^2$, Leo Radzihovsky$^2$, Ana Maria Rey$^1$}
\affiliation{ $^1$JILA, University of Colorado and National Institute of Standard and Technology, Boulder, CO 80309-0440,USA}
\affiliation{ $^2$Department of Physics, University of Colorado, Boulder, Colorado 80309, USA}

\begin{abstract}
We study the resonant effects produced when a Feshbach dimer crosses a scattering continuum band of atoms in an optical lattice.
We numerically obtain the exact spectrum of two particles in a one-dimensional lattice and develop an effective atom-dimer Hamiltonian that accurately captures resonant effects.  The lattice-induced resonances lead to the formation of bound states simultaneously above and below the scattering continuum and significantly modify the curvature of the dimer dispersion relation.
The nature of the atom-dimer coupling depends strongly on the parity of the dimer state leading to a novel coupling in the case of negative parity dimers.

\end{abstract}

\maketitle

Feshbach resonances~\cite{kohler2006production} and optical lattices~\cite{greiner2002quantum}
are two powerful experimental tools used to drive ultracold atomic gases into the strongly correlated many-body regime.
While recent experiments have realized such resonant lattice systems~\cite{PhysRevLett.94.080403}, current understanding of them is limited, with traditional single-band Hubbard model failing to capture the resonant regime.
However, a many-band description  necessary near a resonance (where interaction is much larger than the band width and gap), when implemented directly may be unnecessarily complicated and impractical for analytical studies.
Thus, the development of better suited effective models is highly desired and important first steps have been taken in this direction~\cite{DuanPRL05,DienerPRL06}.

At the two-body level, the lattice potential changes the scattering properties, shifting existing resonances and inducing new ones. This occur when dimers formed with atoms in excited bands cross and hybridize with the scattering continuum of the lowest band~\cite{kestner2010anharmonicity}. While resembling confinement-induced resonances~\cite{Olshanii98}, lattice resonances exhibit richer phenomenology and are more challenging to describe due to the coupling between the relative and the center of mass
degrees of freedom.
Previous two-body lattice studies~\cite{PhysRevA.77.021601} have mainly focused on the lowest molecular state and the lowest atomic band with the virtual higher renormalizing the parameters of the single-band Hubbard model.

\begin{figure}[h]
\begin{center}
\begin{tabular}{cc}
\includegraphics[scale=0.55,angle=0]{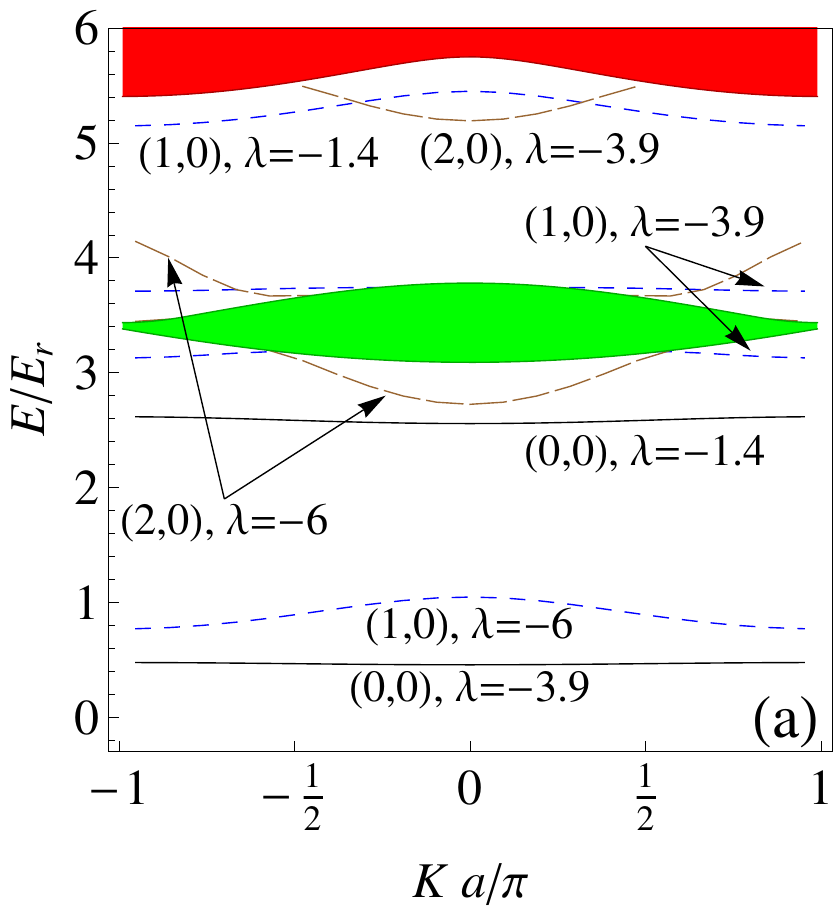}&
\includegraphics[scale=0.5,angle=0]{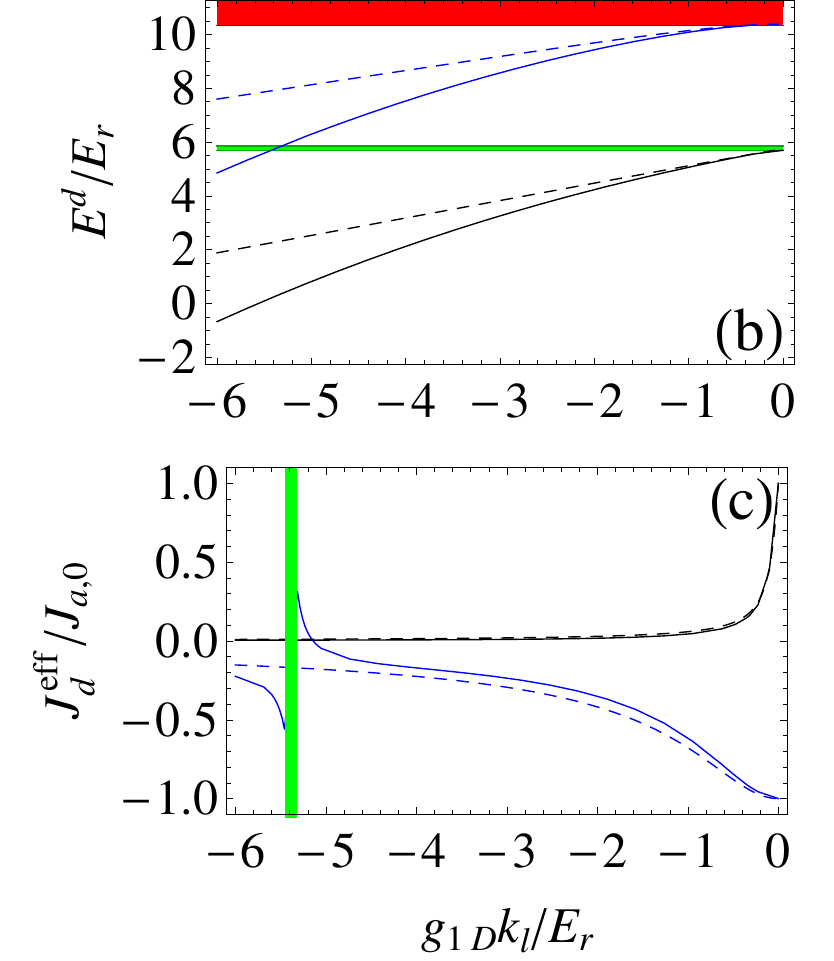}
\end{tabular}
\caption{(Color online) (a) Two-body spectrum for a lattice with $V_0=4 E_r$ as a function of $K$. Colored regions correspond to the scattering continuum bands [lower region corresponds to  $E^{2a}_{0,0}$ (green online), and the upper region corresponds to $E^{2a}_{1,0}$ (red online)].
The curves are the $(0,0)$, $(1,0)$ and $(2,0)$ bound states at different interaction strengths $\lambda$, On the lattice, the (2,0) dimer state is lower in energy than the (1,1) dimer state. The $(2,0)$ dimer state is lower in energy than the (1,1) dimer state.
 (b) and (c) Exact solutions (solid curves) and two-band Hubbard model predictions (dashed curves) of the energy $E^{d}$ (b) and effective tunneling $J^{eff}$ (c) for the lowest two bound states of a lattice of $V=10E_r$. Shaded regions in (b) correspond to $E^{2a}_{0,0}$ and $E^{2a}_{1,0}$ and the shaded region in (c) to the interaction regime in which the excited dimer enters $E^{2a}_{0,0}$.  }
\label{Bands2ANI}
\end{center}
\end{figure}

In this Letter, we study the 1D lattice-induced resonances. We derive an effective atom-dimer many-body model that  captures all the features of our exact numerical solution of the two-body problem summarized in Figs.~\ref{Bands2ANI} and \ref{SpectRes}. These include: (i) attractive and repulsive resonance-induced bound states, that for a range of the center-of-mass momenta can simultaneously appear above and below the two-particle continuum bands, (ii) 1D lattice 
 resonances (absent in 1D lattice-free systems) at the center-of-mass momenta, where a bound state enters the two-particle continuum,
(iii) strong interaction-dependence of the molecular bound-state dispersion, that can even be driven negative.
  Our two-body results are also applicable to two-species fermions in a singlet state. However, for many-system, in this Letter we focus on bosonic atoms. The effective Hamiltonian, inspired by Ref.~\cite{DuanPRL05}, is a lattice projection of a two-channel model~\cite{friedberg1989gap}.
  The novel feature of our model is a nearest neighbor atom ($a_i$) - dimer ($d_i$) Feshbach coupling, 
 \begin{equation}
H_{ad}=g \sum_i d^\dagger_{i}a_{i}(a_{i+1}+(-1)^{P_d} a_{i-1})+\mbox{H.c.}+...,
\label{EffCoupling}
\end{equation}
that depends sensitively on the dimer's parity $P_d=0$ (symmetric) $P_d=1$ (antisymmetric), that is crucial to capturing the dispersion of lattice-induced resonances and bound states. We emphasize that the key atom-dimer coupling $H_{ad}$, Eq.~\eqref{EffCoupling}  proposed here, describes two atoms forming a dimer via a process which depends on the dimer's momentum, thus is not related to the usual two-channel model of Feshbach resonances. We derived Eq.~\eqref{EffCoupling}   via a careful analysis of the dimer orbital structure, supported by exact diagonalization of a two-atom Hamiltonian [Eq.~\eqref{H} below] in a periodic potential.

Our starting point is the bosonic 1D lattice Hamiltonian with contact interaction,
\begin{equation}
H=\sum_\alpha \left(-\frac{\hbar^2}{2m}\frac{\partial^2}{\partial x_\alpha^2}+V_0 \sin^2(k_l x_\alpha)\right)+\sum_{\alpha<\beta}g_{1D} \delta(x_\alpha-x_\beta),
\label{H}
\end{equation}
where $a$ is the lattice constant, $k_l=\pi/a$, and $V_0$ is the lattice depth. The Hamiltonian (2) can be realized with current lattice experiments [10] with tight confinement $a_\perp\ll a_{||}$ ( $a_\perp,_{||}$ are  oscillator lengths lattice wells  in the $\perp$ and $||$ directions) and negligible tunneling in the perpendicular direction ($J_\perp\approx 0$). The interactions are controlled by tuning the scattering length $a_s$ close to a Feshbach or confinement-induced resonance~\cite{Olshanii98}.
 In the regime $|a_s|<a_\perp$, the atoms remain well localized in the transversal ground state and the low-energy behavior is described by Eq.~\eqref{H} with $g_{1D}\approx 4\hbar^2 a_s/(m a_\perp^2)$. The typical lattice energy scale is $E_R=\hbar^2k_l^2/2m$ and the dimensionless interaction strength is $\lambda=g_{1D}k_l/E_r$. In free 1D space ($V_0=0$), the two-body Hamiltonian supports a bound state at an arbitrarily weak interaction with a binding energy $E_b=g_{1D}^2 m/(4\hbar^2)$.

The two-boson Schr\"odinger equation
 is numerically solved for a system of $L$ sites and periodic boundary conditions using a plane wave expansion~\cite{vonstechinprep}. The convergence has been extensively tested by changing $L=1,..., 21$, and the basis dimension. Figures~\ref{Bands2ANI} and~\ref{SpectRes} summarizes the numerical results for different lattice depths and attractive interaction values. States outside
the scattering continuum bands [colored
regions in Figs.~\ref{Bands2ANI}~(a) and~\ref{SpectRes}] represent bound states, while the states
inside the continuum bands are either scattering states or dimer
states which decoupled from the scattering continuum.

 The single particle energies are grouped in to bands  $E^{1a}_n(k)$ with $n\ge0$ which in the tight binding regime take the form $E^{1a}_n(k)=\epsilon_{a,n}-2 J_{a,n}\cos(ka)$, where $\epsilon_{a,n}=\int |w_{n,i}(x)|^2 H_0(x)  dx$ is the onsite energy and $J_{a,n}=\int w^*_{n,i+1}(x) H_0(x) w_{n,i}(x)dx$  is the nearest neighbor tunneling. Here, $H_0$ is the noninteracting part of the $H$~\ref{H}, and $w_{n,i}(x)$ the wannier function of band $n$ centered at site $i$~\cite{jaksch1998cold}.
 The two-body solutions describe the scattering continuum bands  [colored regions in Fig.~\ref{Bands2ANI}~(a)] which are the symmetrized product of the single-particle eigenstates with energies $E^{2a}_{n,m}(K,k)=E^{1a}_n(K/2+k)+E^{1a}_m(K/2-k)$, and can be classified by the
 band label pair $(n,m)$. For each scattering continuum, we can define the typical band gap as $\hbar\omega_{nm}=\min(|\epsilon_{a,n}+\epsilon_{a,m}-\epsilon_{a,k}-\epsilon_{a,l}|)$ with $\{k,l\}\ne\{n,m\}$.

  For weakly interacting atoms ($|g_{1D}/a_{||}|\ll \hbar\omega_{nm}$)  in the tight-binding regime, a bound state is formed in the vicinity of each isolated (not overlapping) scattering continuum $(n,m)$ with parity $P_d=(-1)^{n+m}$.
 A natural starting point to capture the molecular behavior in Fig.~1 is a tight binding model with
each of the two particles with its own (possibly the
same) hoping $J_n$ and $J_m$, and an
on-site interaction  $U_{nm}=g_{1D}\int |w_{m,i}(x)|^2 |w_{n,i}(x)|^2 dx$ (of order of $g_{1D}/a_{||}$). This model can be solved analytically giving a molecular spectrum $E^{d,tb}_{(n,m)}(K)=\epsilon_{a,n}+\epsilon_{a,m}+\mbox{sign}(U_{nm})\sqrt{ U_{nm}^2+4 J_{a,n}^2+4J_{a,m}^2+8J_{a,n} J_{a,m} \cos(K a)}$.
Solid and dashed curves in Figs.~\ref{Bands2ANI}~(b) and (c) present, respectively, the exact and the two-band Hubbard model predictions of the molecular energy $E^{d}=[E^{d,tb}_{(n,m)}(0)+E^{d,tb}_{(n,m)}(\pi/a)]/2$ and the effective molecular hoping $J_d^{eff}=[E^{d,tb}_{(n,m)}(\pi/a)-E^{d,tb}_{(n,m)}(0)]/4$ for the lowest two scattering continua $E^{2a}_{0,0}$
 and $E^{2a}_{1,0}$.
While the two-band description is accurate at weak interaction, it clearly breaks down at larger interaction; it misses molecular band splitting and hybridization with lower 2-particle continuum bands and fails to capture the limit of tightly bound
molecule discussed below.

At stronger interactions ($|g_{1D} /a_{||}|\sim\hbar\omega_{nm}$), an accurate description of the two-body physics requires a number of bands of the order $\mathcal{O} (\hbar\omega_{nm}/J_{a,n})$.
The bound states formed below each two-particle continuum at weak interactions moves downward as the interaction strength increases and eventually cross (and in the process, hybridize with) the lower two-particle continua (see Figs.~\ref{Bands2ANI} (a) and \ref{SpectRes} and~\cite{footnote}).  The specifics of how it hybridizes depends on which band it came from [compare the (1,0) curves at $\lambda=-3.9$  and the (2,0) curves at $\lambda=-6$ in Fig.~\ref{Bands2ANI} (a), and Figs.~\ref{SpectRes}~(a) and (b)] and constitutes the subject of our study. Close to the resonance, the bound state dispersion relation is strongly modified changing the sign $J_d^{eff}$ [see Fig.~\ref{Bands2ANI} (c)]. Figure~\ref{SpectRes} shows the spectrum for the (1,0) and (2,0) lattice resonance deep in the tight-binding regime. States inside $E^{2a}_{0,0}$ allow the determination of the scattering properties of the atoms in the lowest band. Also, resonant effects lead to the appearance of new bound states at either the edges [Fig.~\ref{SpectRes} (a)] or the center [Fig.~\ref{SpectRes} (b)] of the Brillouin zone (BZ). As shown below, the qualitative differences between the (1,0) and (2,0) lattice resonances are accurately captured by the $K$-dependence of the atom-dimer coupling.

Finally, for sufficiently strong interactions ($|g_{1D} /a_{||}|\gg\hbar\omega_{nm}$), the
tightly bound dimers which are not in resonance with the scattering continuum bands are well described by a particle with mass $2m$ moving in a periodic potential with
depth $2V_0$. Our numerical calculations reproduce this limiting behavior.

To gain more physical intuition on the lattice resonance phenomena, we adopt a bosonic variant of an effective two-channel lattice Hamiltonian~\cite{DuanPRL05}. For simplicity, we focus on the resonant two-body physics in an energy
 window around the $(n_0,n_0)$ two-particle continuum. More general $(n_0,m_0)$  continuum bands can also be analyzed in a similar manner.
We consider the interaction regime in which the dimer is close in energy to the  $(n_0,n_0)$ atomic continuum band, i.e., $|\epsilon_d-2\epsilon_{a,n_0}|\ll|\epsilon_{a,n_0+1}-\epsilon_{a,n_0}|$
 where $\epsilon_d$ is the onsite dimer energy. The effective description
 explicitly introduces a localized bare dimer state  $|d_i\rangle=d_i^\dagger|0\rangle$  as the only energetically allowed double occupation of a lattice site. The energies of $n$-occupied sites ($n>2$) are significantly modified by the strong interactions and are expected to fall outside the effective description energy window.
 Thus, we expect that the $n$-occupied  sites are energetically suppressed and we impose this assumption by treating the atoms and dimers as hard-core objects.  The exact many-body Hamiltonian can be formally reduced to the Hilbert space of empty, singly and doubly occupied states~\cite{DuanPRL05} by applying the projector operator $\mathcal{P}_{ad}=\bigotimes_i \mathcal{P}_{ad,i}$ where $\mathcal{P}_{ad,i}=|0_i\rangle \langle  0_i|+|a_i\rangle \langle  a_i|+|d_i\rangle \langle  d_i|$ (here $ a_i$ is the atomic annihilation operator).
The leading interactions terms in the tight-binding regime come from nearest neighbor couplings  which involve tunneling of only one atom: i) the coupling of two atoms to a dimer and ii) the exchange of an atom and a dimer on nearest neighbor sites. Keeping only these terms, the effective Hamiltonian reduces to
\begin{multline}
H_{eff}=\mathcal{P}^\dagger_{ad}\left(\sum _i\epsilon_d d^\dagger_i d_i-J_{d} \sum_{\langle i,j\rangle}  d^\dagger_i d_{j} -J_a \sum_{\langle i,j\rangle}a^\dagger_{i} a_{j} \right.\\ \left.+ g_{ex}\sum_{\langle i,j\rangle}   d^\dagger_{i}d_{j} a^\dagger_{j}a_{i}+\sum_{\langle i,j\rangle} g_{i-j}[ d^\dagger_{i}a_{j}a_{i}+a^\dagger_{i}a^\dagger_{j}d_{i}]\right)\mathcal{P}_{ad},
\label{EffHam}
\end{multline}
where $J_a=J_{a,n_0}$ ($J_d$) is the atomic (dimer) tunneling, $g_{\pm1}$ is the atom-dimer coupling and $g_{ex}$ is the atom-dimer exchange coupling. The Hamiltonian~\eqref{EffHam} can be equivalently understood
as arising from the lattice projection of the two-channel model, restricted to nearest neighbor couplings
between the dominant near resonant closed-channel molecule and nearby two-atom continuum.

\begin{figure}[h]
\begin{center}
\includegraphics[scale=0.7,angle=0]{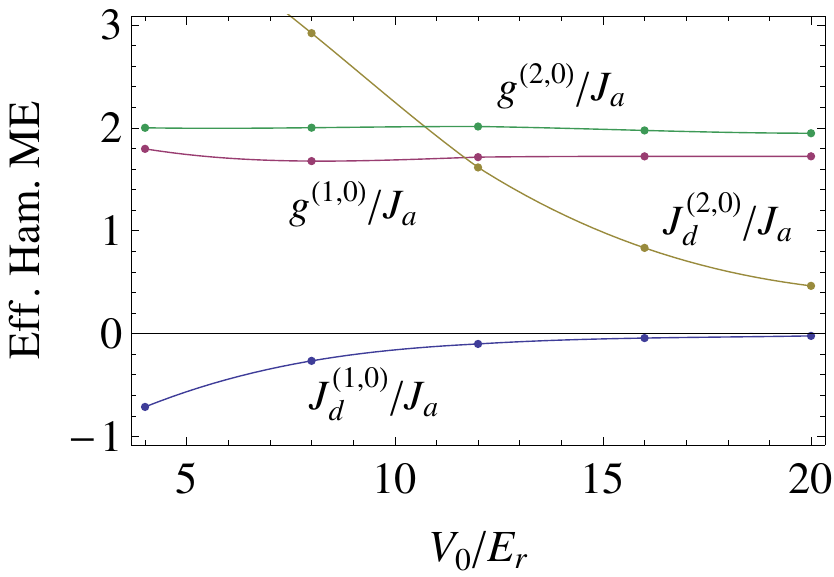}
\caption{(Color online) Effective Hamiltonian matrix elements (ME) as a function of $V_0$.}
\label{EffHamPa}
\end{center}
\end{figure}

A crucial aspect of the proposed effective Hamiltonian is the dependence of the atom-dimer coupling on the relative position ($i-j=\pm1$) between atoms and dimers, $g_{-1}=(-1)^{P_d}g_{+1}$. This dependence comes from the parity properties of dimer Wannier function which are directly related to dimer symmetry with respect to the center-of-mass coordinate. Since the one- and two-body Hamiltonians are invariant under a reflection $\mathcal{R}$ that takes $\{r_i\}\to\{-r_i\}$, the atom and dimer Wannier functions are either symmetric ($s$-orbital) or antisymmetric ($p$-orbital) under the reflection at the bottom of the lattice site in which they are centered.  By applying the reflection operation $\mathcal{R}$ to $g_{\pm1}$, and using parity properties of the Wannier functions we obtain $g_{-1}=(-1)^{P_d}g_{+1}$. This structure is crucial to capture the hybridization of different excited dimers with the two-atom continua (see Fig.~\ref{SpectRes}).

We next turn to the quantitative determination of $H_{eff}$ parameters.
This requires a
full lattice solution, not just single site approximation~\cite{DuanPRL05}.
Associated to a dimer state, there is a dimer Wannier function $W_m(r_1,r_2)$ that can be used to obtain the parameters of the effective Hamiltonian. The dimer Wannier function can be determined by first solving the two-body problem in a reduced Hamiltonian $H'=\mathcal{P}^\dagger H\mathcal{P}$ and then using the dimer Bloch functions $\phi_{m,K}(r_1,r_{2})$ to construct the dimer Wannier function. Here $\mathcal{P}$ projects the Hilbert space outside the $(n_0,n_0)$  scattering continuum band, ensuring the orthogonality between the dimer and scattering states of the $n_0$ band [$\mathcal{P}=1-\mathcal{P}_{2a}$  with $\mathcal{P}_{2a}\equiv\sum_{i\ne j} |a_i a_j\rangle \langle a_i a_j|$].
A similar procedure has been used to extract the ``closed channel'' dimer in confinement induced resonances~\cite{bergeman2003atom}.

The reduced two-body Hamiltonian describes all dimer bands and scattering continuum bands different from ($n_0$,$n_0$).
With the dimer Wannier function (center at site $j$) $W_{m,j}(r_1,r_{2})=W_{m}(r_1-ja,r_{2}-ja)=1/\sqrt{L}\sum_K e^{\mbox{i} K j a}\phi_{m,K}(r_1,r_{2})$ in hand, we extract the two-body parameters of the effective Hamiltonian by requiring that the matrix elements of the $H_{eff}$ match those of the exact Hamiltonian, i.e., $J_{d}=-\langle d_j|H|d^\dagger_{j+1}\rangle$, $\epsilon_{d}=-\langle d_j|H|d^\dagger_j\rangle$ and $g_{\pm1}=\langle d_j|H|a^\dagger_ja^\dagger_{j\pm1}\rangle$.
 Alternatively, $J_d$ and  $\epsilon_{d}$ can be estimated from the dimer band assuming that the dimer dispersion relation follows the tight binding prediction $\epsilon_d(K)=\epsilon_d-2J_d \cos(Ka)$.
 The $g_{ex}$ coupling can be obtained similarly and is expected to be of the order of the $J_a$,
however it only plays a role in a systems with more than two atoms.
Consequently, we leave its quantitative determination to future publications~\cite{vonstechinprep}.

In our numerical implementation, it is not simple to reduce the exact Hamiltonian to $H'$, so we apply an alternative method to extract the dimer Bloch functions. First, we solve the exact Hamiltonian at an interaction for which the dimer is close to the scattering continuum but still weakly coupled to it. Then, we Gram-Schmidt orthogonalize the dimer state from the {\it uncoupled} scattering continuum (described by hard core bosons) to obtain approximate description of the {\it bare} dimer Bloch function and dispersion relation. Finally, we construct the Wannier functions and extract the effective Hamiltonian parameters.
 These parameters, illustrated in Fig.~\ref{EffHamPa}, converge fast with the lattice size and only a few sites are needed to reach convergence.

\begin{figure}[h]
\begin{center}
\begin{tabular}{ccc}
\includegraphics[scale=0.28,angle=0]{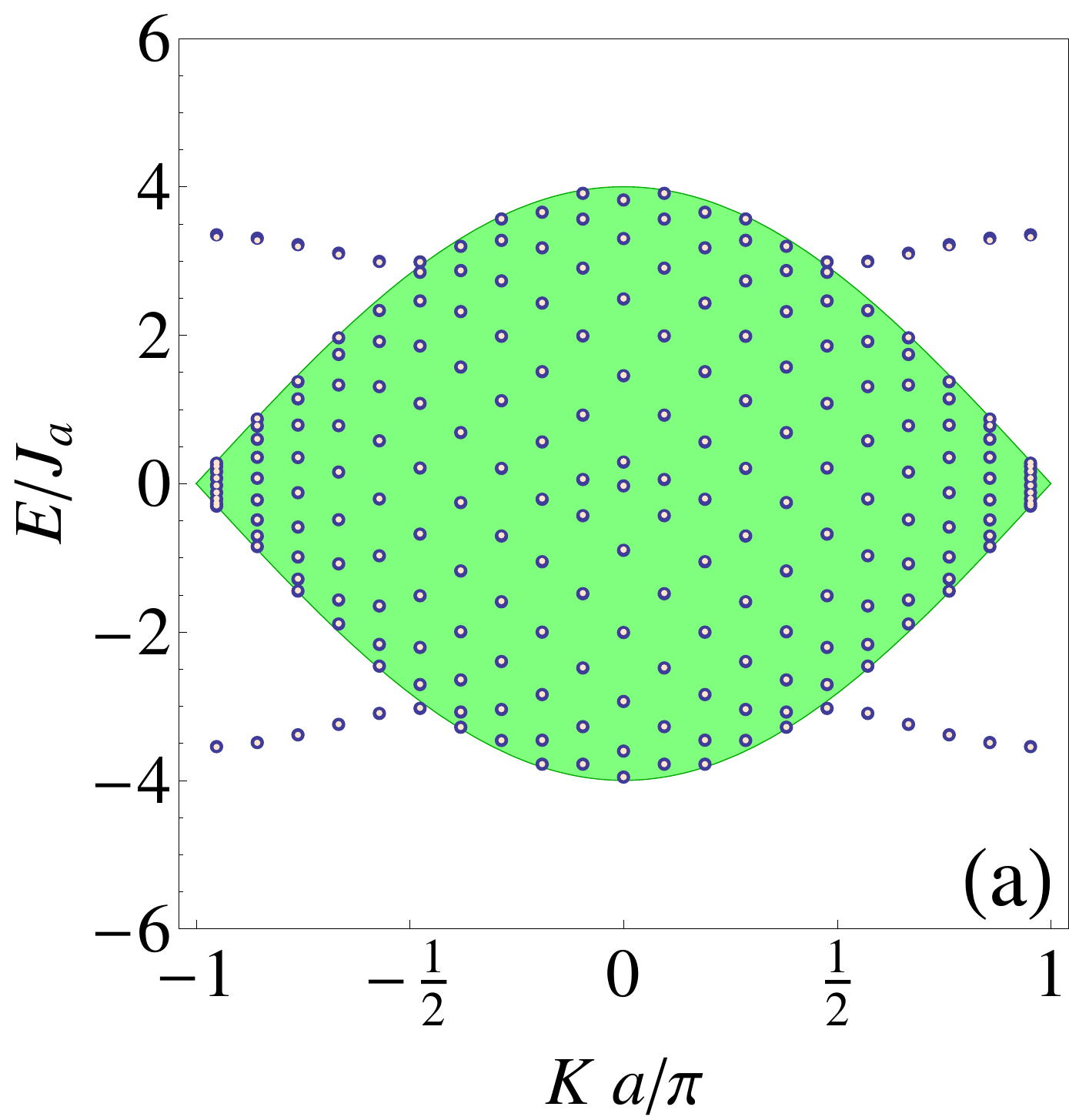}&\,\,\,\,\,&
\includegraphics[scale=0.28,angle=0]{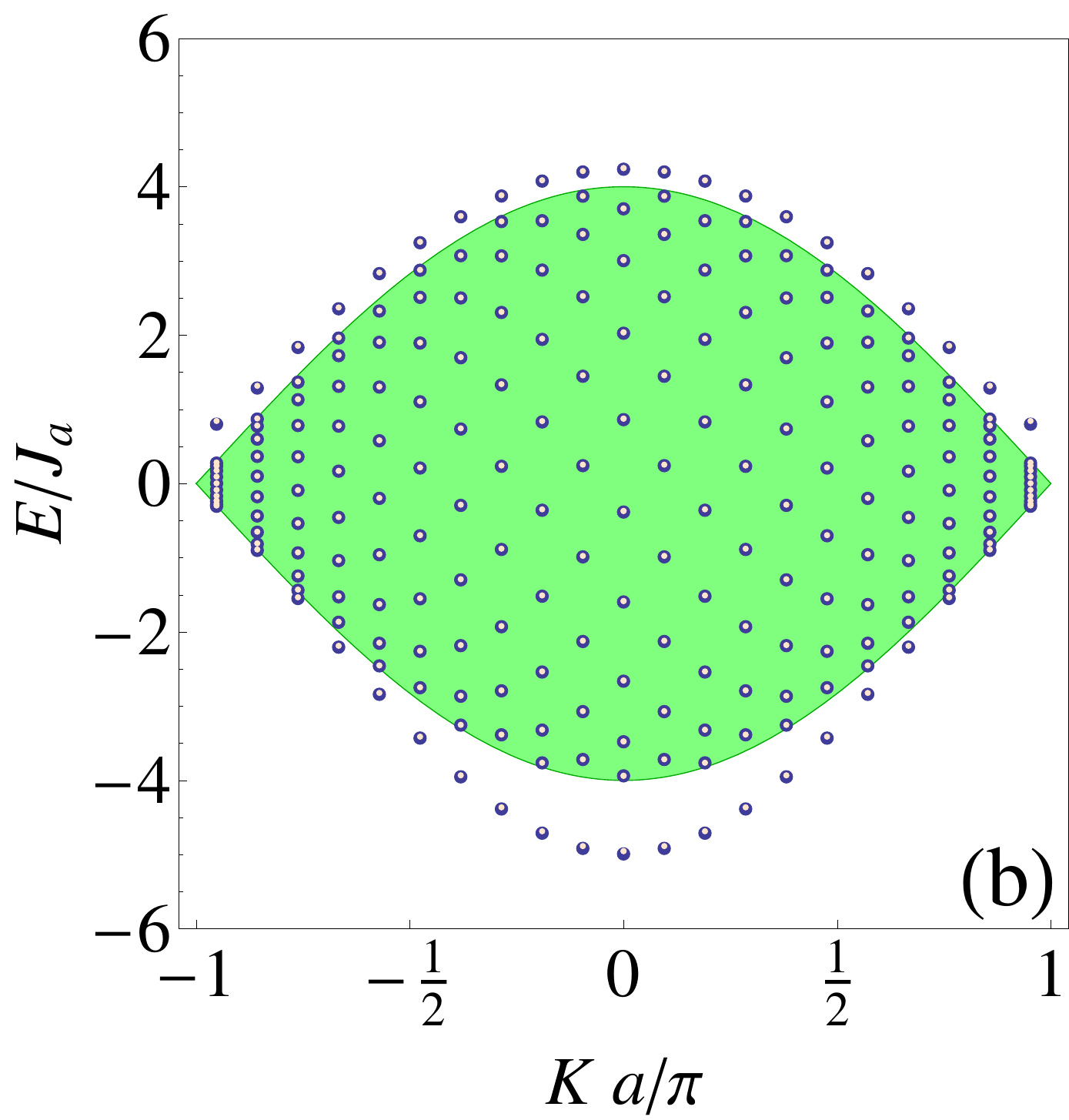}
\end{tabular}
\caption{(Color online) Two-body spectrum at the (1,0) (a) and (2,0) (b) lattice resonances. Comparison of the exact and the model Hamiltonian predictions for a lattice of 21 sites with periodic boundary conditions. Calculations for $V_0=20 E_r$ at interactions for which the (1,0) and (2,0) dimers are in resonance with the (0,0) scattering continuum. Shaded region corresponds to the (0,0) scattering continuum. Larger and darker  (blue online) symbols correspond to the exact Hamiltonian predictions while smaller and brighter (orange online) correspond to the model predictions. The two results are almost indistinguishable. }
\label{SpectRes}
\end{center}
\end{figure}
The bound state and scattering two-body solutions of the effective Hamiltonian  \eqref{EffHam},\eqref{EffCoupling} can be solved analytically for an infinite lattice.
When solving the two-body problem at fixed $K$, the atom-dimer coupling is proportional to $g_S(K)=2g \cos(Ka/2)$ for $P_m=0$ and $g_A(K)=2g\mbox{i} \sin(Ka/2)$ for $P_m=1$. Thus, the atom-dimer coupling is maximum at $K=0$ for
$s$-$s$ coupling but at $K=\pi/a$ for $s$-$p$  coupling. The two-body bound state energies are given by $E=J(K)(\alpha+1/\alpha)$ where $J(K)=2J\cos(ka/2)$ and $\alpha$ is the solution of $J(K)-\epsilon_d(K)\alpha+(|g_\beta(K)|^2-J^2(K))\alpha^2/J(K)=0$ with the constraint $|\alpha|<1$. The analytical solutions allow from zero to two bound states as found in the exact numerical solution illustrated in Figs.~\ref{Bands2ANI},~\ref{SpectRes}. A second bound state appear for the $K$ values at which the coupling is maximum, i.e, edges of the BZ for $P_m=1$ [see e.g. Fig.~\ref{SpectRes}~(a)]  and center of the BZ for $P_m=0$ [see e.g. Fig.~\ref{SpectRes}~(b)]. If the dimer is inside the scattering continuum and the coupling is weak, the system supports metastable  states whose lifetime can be obtained by analytically continuing the bound state solutions into the complex plane [see e.g. small $K$ region in Fig.~\ref{SpectRes}~(a)]. The analytical solutions also reproduce the two-body bound states of the single-band Hubbard model~\cite{winkler2006repulsively} when
$|d_i\rangle=a^\dagger_i a^\dagger_i |0\rangle/\sqrt{2}$ and the coupling is of the symmetric parity  type (i.e., $g_S(K)$). Under these conditions, $g=\sqrt{2}J_a$, $\epsilon_d$ is the onsite interaction energy $U$, $J_d=0$, and the bound state energies are $E=\mbox{sign}(U)\sqrt{U^2+16J^2 \cos^2(ka/2)}$. As shown in Fig.~\ref{EffHamPa}, the atom-dimer coupling is even stronger for excited dimers. For $V_0=40 E_r$, $g^{(1,0)}\approx 1.72 J_a$ and  $g^{(2,0)}\approx 1.90 J_a$ and these values only change by a few percents in the range $V_0=4$--$40 E_r$ and are in good qualitative agreement with the predictions of Ref.~\cite{kestner2010anharmonicity}.

To confirm the validity of the effective Hamiltonian \eqref{EffCoupling}, \eqref{EffHam},we solve the two-body problem for a finite lattice with periodic
boundary conditions using both $H$ and $H_{eff}$ in the resonant regime. We found excellent agreement between the exact and the effective Hamiltonian predictions [see Figure ~\ref{SpectRes}].

The $s$-$s$ and $s$-$p$ orbital symmetry of the coupling can be experimentally probed by analyzing the quasimomentum dependence of the molecules fraction after a magnetic field ramp through a lattice induce resonance. Initially dimers are formed in excited bands, which can be achieved by populating atoms in excited bands~\cite{muller2007state}. Then, interactions are tuned through a lattice resonance, and finally, the dimer fraction is measured as a function of the dimer quasimomentum. At the two-body level, the final dimer fraction is well described by a Landau-Zener probability $\exp(-\delta_{LZ})$ with a Landau-Zener parameter $\delta_{LZ}\propto |g_\beta(K)|^2/|\alpha|$ where $\alpha$ is the speed of the ramp. Thus, the final molecule probability will be mainly affected at the center ($s$-$s$) or the edges ($s$-$p$) of the BZ depending on the symmetry of the coupling.

Our predictions on the general structure of the atom-dimer coupling are based only on symmetry properties of the atom and dimer Wannier functions and can be easily extended to multi-component and higher dimensional systems. The structure of the lattice induce resonances in higher dimension would be determined by the dimer symmetry in each lattice axis direction.
 Thus, we expect a rich variety of physical phenomena when the system is close to a lattice-induced resonance, that extends to higher dimensions.

This work was supported by the NSF (PFC
and Grant No. PIF-0904017), the ARO through the DARPA-OLE (J. v. S. and A. M. R.), NSF
grant DMR-1001240 (L. R.), and NSF grant DMR-0449521 (V. G.).


\begin{thebibliography}{10}

\bibitem{kohler2006production}
T. K{\"o}hler {\it et al.}, Rev. Mod. Phys. {\bf 78},  1311
  (2006).

\bibitem{greiner2002quantum}
M. Greiner {\it et al.},  Nature {\bf
  415},  39  (2002).

\bibitem{PhysRevLett.94.080403}
M. K\"ohl {\it et al.},  Phys. Rev.
  Lett. {\bf 94},  080403  (2005).

\bibitem{DuanPRL05}
L.-M. Duan, Phys. Rev. Lett. {\bf 95},  243202  (2005).

\bibitem{DienerPRL06}
R.~B. Diener and T.-L. Ho, Phys. Rev. Lett. {\bf 96},  010402  (2006).
D.~B.~M. Dickerscheid, {\it et al.}, Phys.
  Rev. A {\bf 71},  043604  (2005). K. Hazzard and E. Mueller, Phys. Rev. A {\bf 81},  31602  (2010).


\bibitem{kestner2010anharmonicity}
J. Kestner and L. Duan, New J. of Phys. {\bf 12},  053016  (2010); Phys. Rev. A {\bf 81},  043618  (2010).

\bibitem{Olshanii98}
M. Olshanii, Phys. Rev. Lett. {\bf 81},  938  (1998). D.~S. Petrov, M. Holzmann, and G.~V. Shlyapnikov, Phys. Rev. Lett. {\bf 84},
  2551  (2000). H. Moritz {\it et al.}, Phys. Rev.
  Lett. {\bf 94},  210401  (2005). E. Haller {\it et al.}, Science {\bf 325},  1224  (2009).


\bibitem{PhysRevA.77.021601}
N. Nygaard, R. Piil, and K. M\o{}lmer, Phys. Rev. A {\bf 77},  021601  (2008).
G. Orso {\it et al.}, Phys. Rev. Lett. {\bf  95},  60402  (2005).
P. Fedichev, M. Bijlsma, and P. Zoller, Phys. Rev. Lett. {\bf 92},  80401
  (2004). H.~P. B\"uchler, Phys. Rev. Lett. {\bf 104},  090402  (2010).
X. Cui, Y. Wang, and F. Zhou, Phys. Rev. Lett. {\bf 104},  153201  (2010).


\bibitem{friedberg1989gap}
R. Friedberg and T.~D. Lee, Phys. Rev. B {\bf 40},  6745  (1989).
V. Gurarie and L. Radzihovsky, Annals of Phys. {\bf 322},  2  (2007).

\bibitem{liao2010spin}
Y. Liao {\it et al.}, Nature {\bf 467},  567  (2010).

\bibitem{vonstechinprep}
J. von Stecher {\it et al.}, in preparation.

\bibitem{jaksch1998cold}
D. Jaksch {\it et al.}, Phys. Rev. Lett. {\bf 81},  3108-3111  (1998).

\bibitem{footnote}
An evolution of the spectrum as interactions are tuned can be found at
\href{http://jila.colorado.edu/~arey/research/index1.html}{http://jila.colorado.edu/\textasciitilde arey/research/index1.html}.


\bibitem{bergeman2003atom}
T. Bergeman {\it et al.}, Phys. Rev. Lett. {\bf 91},  163201
  (2003).

\bibitem{winkler2006repulsively}
K. Winkler {\it et al.}, Nature {\bf 441},  853  (2006).

\bibitem{muller2007state}
T. M{\"u}ller {\it et al.}, Phys. Rev. Lett.  {\bf 99},  200405  (2007).

\end{thebibliography}
\end{document}